\documentclass[fleqn,twoside]{article}
\usepackage{espcrc2}
\newcommand{\beq}{\begin{equation}}
\newcommand{\eeq}{\end{equation}}
\newcommand{\beqa}{\begin{eqnarray}}
\newcommand{\eeqa}{\end{eqnarray}}
\newcommand{\beqar}{\begin{eqnarray*}}
\newcommand{\eeqar}{\end{eqnarray*}}

\newcommand{\al}{\alpha}
\newcommand{\be}{\beta}

\def\Tr           {\mbox{\rm Tr}\,}

\def\ran          {\rangle}
\def\lan          {\langle}
\def\fsH    {H\!\!\!\!/\,}

\newcommand{\eps}{\epsilon}

\newcommand{\inn}{\!\cdot\!}

\newcommand{\lam}{\lambda}

\newcommand{\z}{\zeta}

\newcommand{\ie}{{\it i.e.,}\ }
\newcommand{\labell}[1]{\label{#1}} %{\label{#1}} %
\newcommand{\reef}[1]{(\ref{#1})}
\begin{document}

% change this to the following line for use with LaTeX2.09
% \documentstyle[twoside,fleqn,espcrc2]{article}

% if you want to include PostScript figures
%\usepackage{graphicx}
% if you have landscape tables
%\usepackage[figuresright]{rotating}

% put your own definitions here:
%   \newcommand{\cZ}{\cal{Z}}
%   \newtheorem{def}{Definition}[section]
%   ...
\newcommand{\ttbs}{\char'134}
\newcommand{\AmS}{{\protect\the\textfont2
  A\kern-.1667em\lower.5ex\hbox{M}\kern-.125emS}}

% add words to TeX's hyphenation exception list
\hyphenation{author another created financial paper re-commend-ed Post-Script}
% declarations for front matter
%{ CERN-PH-TH/2011-027}

\title{Three Point Tree Level Amplitude in Superstring Theory}
\author{Ehsan Hatefi \address[MCSD]{Theory Group, Physics Department,
         CERN CH-1211, Geneva 23, Switzerland.}%
        \thanks{On leave from Ferdowsi University of Mashhad.
        ehsan.hatefi@cern.ch}}
       
\begin{abstract}
In order to check the definite amplitude and the exact zero result of the amplitude of  three massless points $(CAA)$ in both string theory and field theory side for $p=n$ case and to find all gauge field couplings to R-R closed string, we investigate the disk level S-matrix element of one Ramond-Ramond field and two gauge field vertex operators in the world volume of BPS branes. 
\vspace{1pc}
\end{abstract}

\maketitle

\vspace*{-7cm}
\noindent
{CERN-PH-TH/2011-027}
\vspace*{5cm}

\section{Introduction}

 Due to the fact that D$_p$-branes are the 
source of Ramond-Ramond $(p+1)$-form fields in IIA and IIB string theories \cite{Polchinski:1995mt}, and that many properties of
them have been investigated  \cite{Witten:1995im,Polchinski:1996na}, they are of high importance 
in both theory and phenomenology.
The stable D$_p$-branes ($p$ is even in IIA
 and odd in IIB theory) preserve half of
supersymmetry. 
 
  Stability, supersymmetry, conserved Ramond-Ramond (RR) charge and having no tachyons are,in fact, all properties of these type II
D$_p$-branes. All supersymmetric D$_p$-branes in
IIA can be generated as bound states of D$_9$-branes \cite{hor1}.
They can also be derived
from K-theory~\cite{wit1}. At leading order, the low-energy action for fields 
corresponds to dimensional reduction of a ten-dimensional U(1) super-Yang Mills theory. When derivatives
of the field strengths are small on the string scale, the
action to all orders in the field strength takes the Born-Infeld form \cite{Dbrane,Leigh:1989jq} (also see
\cite{Tseytlin:1999dj}).
The low energy action describing the dynamics of  D$_p$-branes consists of two
parts.
The first part is the Born-Infeld action (for more details see \cite{Myers:1999ps}). In addition to providing the kinetic terms for the world-volume fields, the BI action  contains the couplings of the D$_p$-brane to the massless Neveu-Schwarz fields in the
bulk. 
 The second part is the Wess-Zumino action, which contains the coupling of the U($N$) 
massless world volume vectors to the closed string RR
field \cite{Polchinski:1995mt,li1996a}.
One method for finding these effective actions is BSFT \cite{Kraus:2000nj}. 
To study WZ couplings for BPS branes,  we use the S-matrix method as the second approach. 

 %%%%%%%%%%%%%%%%%%%%%%%%%%%%%%%%%%%%%%%%%%%%%%%%%%%%%%%%%%%%%%%%%%%%%%%%%%%%%%%%%%%%%%%
\section{The Three Point Superstring Scattering (CAA) }
One important tool in string theory is scattering theory \cite{Veneziano:1968yb}. In this section, using the conformal field theory techniques \cite{Kostelecky:1986xg}, we evaluate this scattering amplitude to find all couplings of one closed string RR field to two gauge fields on the world-volume of a single BPS D$_p$-brane with flat empty space background. A great deal of effort for the understanding of scattering amplitudes at tree level has been made
\cite{Hashimoto:1996bf}. Some previous works on scattering including a D$_p$-brane and some other works about
applications on D$_p$-brane can be found in \cite{bachas}.

To calculate a S-matrix element, one needs to choose the picture of the vertex operators  appropriately. The sum of the superghost charge must be -2 for the disk level amplitude.  Hence, this S-matrix element is given by the following correlation function
\begin{eqnarray}
{\cal A}^{AAC} & \sim & \int dx_{1}dx_{2}dzd\bar{z}\,
  \lan V_{A}^{(-1)}{(x_{1})}
V_{A}^{(0)}{(x_{2})}
\nonumber\\&&\times
V_{RR}^{(-1/2,-1/2)}(z,\bar{z})\ran,\labell{sfield}\eeqa

where the vertex operator and the ``doubling trick'' have been mentioned in \cite{Hatefi:2010ik}.  The Wick-like rule~\cite{Liu:2001qa} is used to find the correlator of $\psi$.
The only subtlety in the use of  this formula  for currents is that the Wick-like contraction for the two $\psi$'s in one current \cite{Garousi:2008ge} should not be considered, so that it follows, 
\beqa
{\cal A}^{CAA} \sim \int dx_{1}dx_{2} dx_{3}dx_{4}(P_{-}\fsH_{(n)}M_p)^{\al\be}I\xi_{1a}\nonumber\cr
\times \xi_{2b}x_{34}^{-1/4}(x_{23}x_{24})^{-1/2}\left(a^a_1a^b_2+2ik_{1c}I^{bac}\right),\nonumber\eeqa
where  
\beqa
I&=&|x_{12}|^{-2s}|x_{13}x_{14}|^{s}|x_{23}x_{24}|^{s}|x_{34}|^{-2s}
\nonumber\\
a^a_1&=&ik_2^{a}\bigg(\frac{x_{32}}{x_{13}x_{12}}+\frac{x_{42}}{x_{14}x_{12}}\bigg)\nonumber\\
a^b_2&=&2^{-1/2}x_{34}^{-3/4}(x_{23}x_{24})^{-1/2}(\gamma^{b}C^{-1})_{\alpha\beta}.
\nonumber\eeqa
For details about  $I^{bac}$ see \cite{Hatefi:2010ik}.  We define the Mandelstam variable as $s=-\frac{\alpha'}{2}(k_1+k_2)^2$. Clearly it is seen 
that the integrand is invariant under
SL(2,R) transformation. Gauge fixing  this  symmetry by fixing the position of the closed string vertex operator as 
 \beqar
 x_{1}&=&x ,\qquad x_{2}=-x,\qquad x_{3}=i,\qquad x_{4}=-i,
 \eeqar
%Note that Jacobian now is $2i(1+x^2)$ .
 the amplitude becomes :
 \beqa
{\cal A}={\cal A}_{1}+{\cal A}_{2}+{\cal A}_{3},
\eeqa

where 
\beqa
{\cal A}_{1}&\!\!\!\sim\!\!\!&ik_{2}^{a}(2i)^{-2s}2^{-1/2}\xi_{1a}\xi_{2b}
\Tr(P_{-}\fsH_{(n)}M_p\gamma^{b})\nonumber\\&&\times
\int_{0}^{\infty} dx
\frac{1}{x}(x^2-1)(2x)^{-2s}(x^2+1)^{2s-1}\nonumber\eeqa
\beqa
{\cal A}_{2}&\!\!\!\sim\!\!\!&-k_{1c}(2i)^{-2s}2^{1/2}\xi_{1a}\xi_{2b}
\Tr(P_{-}\fsH_{(n)}M_p\Gamma^{bac})
\nonumber\\&&\times
\int_{0}^{\infty} dx(2x)^{-2s}(x^2+1)^{2s-1}%\Gamma^{lj}
 %\nonumber\\&&\times(\delta_{p,n+1}+\delta_{p,n-1}+\delta_{p,n+2})
\eeqa
\beqa
{\cal A}_{3}&\!\!\!\sim\!\!\!&-k_{1c}(2i)^{-2s}2^{1/2}\xi_{1a}\xi_{2b}
(\eta^{cb}\Tr(P_{-}\fsH_{(n)}M_p\gamma^{a})\nonumber\\&&-\eta^{ab}\Tr(P_{-}\fsH_{(n)}M_p\gamma^{c}))\nonumber\\&&\times\int_{0}^{\infty} dx
\frac{1}{2ix}(-x^2+1)(2x)^{-2s}(x^2+1)^{2s-1}.
 \nonumber\eeqa
To obtain the definite amplitude, the real part of $s$ has to be less than zero.
In addition all integrations must be taken over the positive values of $x$, otherwise 
the terms $x^{-s}$ will give rise to a complex amplitude.
Thus, the only non vanishing integral is the second one for which the result is  
\beqa
\int_{0}^{\infty} dx(2x)^{-2s}(x^2+1)^{2s-1}&=&\frac{\pi^{1/2}\Gamma[-s+1/2]}{2 \Gamma[-s+1]}.\nonumber\eeqa

 To compare the field theory which apparently has  massless field, \ie  WZ action,  with the above amplitude, it must be  expanded such that the  massless pole of the field theory survives  and all other poles disappear in the form of contact terms. 
Note that the S-matrix element of all four point massless vertex operators in superstring theory is also
 found in standard books \cite{mgjs,jp}.

\section{Momentum expansion}
 
Our goal is  to examine the limit of $\alpha'\rightarrow 0$  of the above string amplitude. Applying  momentum conservation along the world volume of the brane, we see that the Mandelstam variable satisfies
the constraint 
\beqa
s=-p_ap^a/2.
\labell{cons}\eeqa
 It has been shown in \cite{Hatefi:2010ik} that the momentum expansion of a S-matrix element should be in general around  $(k_i+k_j)^2\rightarrow 0$ and/or $k_i\inn k_j\rightarrow 0$.  The amplitude must only have a massless pole in the $(k_1+k_2)^2$channel, so that the correct momentum expansion at the low energy limit for s-channel must be around
$(k_1+k_2)^2\rightarrow 0$. Using the on-shell relations, they can be rewritten in terms of the Mandelstam variable as
$
s\rightarrow 0$. 
Under the constraint \reef{cons}, note that $p_ap^a\rightarrow 0$  is allowed for D-branes. Therefore the S-matrix
element can be evaluated for BPS branes. Thus, expansion of the functions 
around the above point will be
\beqa
(2)^{-2s}\frac{\pi^{1/2}\Gamma[-s+1/2]}{\Gamma[-s+1]}&=&{\pi}\left( \sum_{n=-1}^{\infty}b_n(s)^{n+1}\right).
\nonumber\eeqa
where some of the coefficients $b_n$ are
\beqa 
&&b_{-1}=1,\,b_0=0,\,b_1=\frac{1}{6}\pi^2,\,b_2=2\z(3).\nonumber\eeqa
As expected, it is seen that the obtained coefficients $b_n$ and the coefficients appearing in the momentum expansion of the S-matrix element of one RR, two gauge fields and one tachyon vertex operator \cite{Garousi:2007fk} are exactly the same.
\section{Low Energy Field Theory }
%%%%%%%%%%%%%%%%%%%%%%%%%%%%
%%%%%%%%%%%%%%%%%%%%%%
%In the S-matrix element \reef{sstring} the external states are only three gauge fields and a closed string Ramond-Ramond field
%in which the gauge fields appear as on-shell or off-shell state. Therefore 
We focused on the part of effective field theory of
D-branes which includes only gauge fields. It is possible to extract the necessary terms from the covariant Born-Infeld action
constructed as the effective D-brane action. The Born-Infeld action is an action for all orders of $\alpha'$. As low energy non-abelian extension of the action, the symmetrized trace of non-abelian generalization of Born-Infeld action was proposed. The non-abelian field strength and covariant
derivative of the gauge field are defined, respectively, as
 \beqa F^{ab}=
\partial^aA^b-\partial^bA^a-i[A^a,A^b],\\
D_aF_{bc}=\partial_aF_{bc}-i[A_a,F_{bc}]. \nonumber\eeqa
%In the field theory \reef{qt1}, $A_a$ is in the adjoint
%representation of the gauge symmetry $U(N)$, where $N$ is the
%number of D-branes. That is
where $A_a=A_a^{\alpha}\Lambda_{\alpha}$
and $\Lambda_{\alpha}$ are the hermitian matrices.
%$\Lambda_{\alpha}$ are in the adjoint representation of the $U(N)$
%group.
 Our conventions for $\Lambda^{\alpha}$ are
\beqa
\sum_\alpha
\Lambda^{\alpha}_{ij}\Lambda^{\alpha}_{kl}=\delta_{ik}\delta_{jl}\,\,,\,\,~~~
\Tr(\Lambda^{\alpha}\Lambda^{\beta})&\!\!\!=\!\!\!&\delta^{\alpha\beta}.
\nonumber\eeqa
%The term of the above expansion which has
%contribution to the above  S-matrix element is in the following  
%\beqa {\cal
%L}&=&-T_p(\pi\alpha')\Tr\left(-(\pi\alpha')F_{ab}F^{ba}\right)\labell{expandL}\\
%\nonumber\eeqa

%%%%%%%%%%%%%%%%%%%%%%%%%%%%%%%%%%%%%%%%%%%%%%%%%%%%%%%%%%%%%%%%%%%%%%%%%%%%%%%%%%%%%%
\section{$p=n+2$ case}

Due to the fact that only  ${\cal A}_2$ is non-zero,  we are not interested in fixing the overall sign of the amplitudes.
 %(since they are
%unimportant for our purposes). Hence in the rest of the equations in this paper we have payed no attention to the sign of  the amplitudes. 
Taking into account the related trace thus the string amplitude should be
\beqa
{\cal A}^{CAA}=\pm(\mu_p \pi^{1/2})\frac{32}{(p-2)!}k_{1c}\xi_{1a}\xi_{2b}
\eps^{baca_{1}\cdots a_{p-3}}\nonumber\\\times 
H_{a_{1}\cdots a_{p-3}}(2)^{-2s}\frac{\pi^{1/2}\Gamma[-s+1/2]}{ 2\Gamma[-s+1]},\label{111}
\nonumber\eeqa
where we normalized the amplitude by $ (\mu_p 2^{1/2} \pi^{1/2})$.
This amplitude is zero upon interchanging gauge fields, rendering the whole amplitude is zero for an abelian gauge group. The amplitude also satisfies the Ward identity. Since the Gamma function has no tachyon/massless pole, the amplitude only has contact terms. The leading contact term is reproduced by the coupling 
\beqa
\frac{1}{2!}\mu_p(2\pi\alpha')^{2}\Tr (C_{p-3}\wedge F\wedge F).\labell{hderv}
\eeqa
The non-leading order terms have to correspond to the higher derivative extension of the above coupling. Thus, the higher vertex will be
\beqa
V(C_{p-3},A_3,A)=\frac{\mu_p(2\pi\alpha')^2}{(p-2)!}\eps^{a_{1}\cdots a_{p+1}}H_{a_{1}\cdots a_{p-2}}\nonumber\\\times\xi_{1a_{p-1}}k_{1a_{p}}\xi_{2a_{p+1}}\sum_{n=-1}^{\infty}b_n(\alpha'k_1.k_2)^{n+1}.\nonumber\\
\nonumber\eeqa

\section{$p=n$ case}

Despite the fact that in string theory both ${\cal A}_1 ,{\cal A}_3 $ are zero in this case, we would like to perform the field theory calculations 
to confirm that, there is no compensation of the massless pole,
\beqa
{\cal A}&=&V^a_{\alpha}(C_{p-1},A)G^{ab}_{\alpha\beta}(A)V^b_{\beta}(A,A_1,A_2),\nonumber\eeqa
where the vertices and propagator are 
\beqa
V^a_{\alpha}(C_{p-1},A)&=&\frac{i\mu_p(2\pi\alpha')}{(p)!}\eps^{a_{0}\cdots a_{p-1}a}\nonumber\\&&\times
H_{a_{0}\cdots a_{p-1}}\Tr(\Lambda_\alpha),\nonumber\eeqa
\beqa
V^b_{\beta}(A,A_1,A_2)=\bigg[\xi_1^b(k_1-k).\xi_2+
\xi_2^b(k-k_2).\xi_1\nonumber\\+\xi_1.\xi_2(k_2-k_1)^b\bigg](-iT_p(2\pi\alpha')^{2}\Tr(\lam_1\lam_2\Lambda_\beta)),\nonumber\eeqa
\beqa
G_{\alpha\beta}^{ab}(A)=\frac{i\delta_{\alpha\beta}\delta^{ab}}{(2\pi\alpha')^2 T_p
(s)}.\nonumber\eeqa
 $\alpha,\beta$ and $a,b$ are the group and world volume indices, respectively. The propagator is derived from the standard gauge kinetic term arising in the expansion
of the Born-Infeld action. Note that the vertex $V^b_{\beta}(A,A_1,A_2)$ is found from the standard non abelian kinetic term of the gauge field. Also, the vertex $V^a_{\alpha}(C_{p-1},A)$ is found from the WZ coupling $C_{p-1}\wedge F$. In the above formula $k$ is the momentum of the off-shell gauge field. The important point should be made is that the vertex $V^b_{\beta}(A,A_1,A_2)$ has no higher derivative correction as it arises from the kinetic term of the gauge field. This vertex has already been found in \cite{Hatefi:2010ik}. Considering those vertices,  the amplitude yields
\beqa
{\cal A}&=&[\xi_{1a}(k_1-k).\xi_2+\xi_{2a}(k-k_2).\xi_1+\xi_1.\xi_2\nonumber\\&&(k_2-k_1)_{a}]
i\mu_p(2\pi\alpha')\frac{1}{(p)!s}\Tr(\lam_1\lam_2)\nonumber\\&&\times\eps^{a_{0}\cdots a_{p-1}a}H_{a_{0}\cdots a_{p-1}},\label{not pure}\eeqa
which of course describes the apparent massless pole in field theory, while there is no massless pole in string theory.

\section{Remarks}
In the $p=n$ case, there is no massless pole at $s =0$. It can be concluded that the kinematic factor provides a compensating
factor of $s$,  but we do not know how the compensation is achieved.
%Even though we have some contact terms in field theory side 
%while in string theory side we have the exact zero result!! 

To understand the vanishing amplitude, remember that to produce the correct amplitude, we must consider all possible orderings of non abelian gauge fields, which means that we must consider resulting terms by interchanging 1 to 2 in \reef{not pure} as well. Note that the quantity in the square bracket in \reef{not pure} is antisymmetric under interchanging 1 to 2. Therefore, apart from the coefficients, the final result for the amplitude is
given by
\beqa
{\cal A}&=&[\xi_{1a}(k_1-k).\xi_2+\xi_{2a}(k-k_2).\xi_1
+\xi_1.\xi_2\nonumber\\&&\times(k_2-k_1)_{a}]
\eps^{a_{0}\cdots a_{p-1}a}H_{a_{0}\cdots a_{p-1}}\nonumber\\&&\times
(\Tr(\lam_1\lam_2)-\Tr(\lam_2\lam_1)).\nonumber\eeqa

The fact that the amplitude is zero, indicates  two concrete points. First, gauge fixing has to be done over the positive values of $x$; otherwise we will have a complex amplitude. Therefore, the upper and lower bound of the integration in string theory have been chosen correctly. Second, there was an apparent massless pole in field theory. However, the amplitude vanishes not because of compensating Mandelstam variable, but because of considering all orderings of gauge fields. There is no contact term for $p=n$ case.

%%%%%%%%%%%%%%%%%%%%%%%%%%%%%%%%%%%%%%%%%%%%%%%%%%%%%%%%%%%%%%%%%%%%%%%%%%%%%%%%%%%%%%%
\section*{Acknowledgments}
The author would like to thank G.Veneziano for very beneficial and enjoyable collaboration throughout the project. He would also like to thank his advisors Luis \'Alvarez-Gaum\'e and M.R Garousi for several valuable discussions . The author also acknowledges	N. Arkani-hamed, L.Dixon, I.Antoniadis , C.Grojean, P.Vanhove, M. Douglas, N.Lambert, J.Drummond and A. Strominger for comments, useful suggestions and the CERN Theory Group for its hospitality. This work was supported Under the Marie Curie or the EU grant UNILHC PITN-GA-2009-237920 .
%%%%%%%%%%%%%%%%%%%%%%%%%%%%%%%%%%%%%%%%%%%%%%%%%%%%%%%%%%%%%%%%%%%%%%%%%%%%%%%%%%%%%%%

 %Use
%\verb+\cite+ to refer to the entries in the bibliography so that your
%accumulated list corresponds to the citations made in the text body.

\end{document}